\documentclass{ws-mpla}

\begin{document}

\markboth{Huaifan li} {Mapping Hawking temperature in the spinning
constant curvature black hole spaces  into Unruh temperature}

\catchline{}{}{}{}{}

\title{Mapping Hawking temperature in the spinning constant curvature black hole spaces  into Unruh temperature}

\author{\footnotesize Huaifan Li}

\address{Department of Physics and Institute of
Theoretical Physics, Shanxi Datong University\\ Datong 037009,
China\\
Key Laboratory of Frontiers in Theoretical Physics, Institute of
Theoretical Physics, Chinese Academy of Sciences, P.O. Box 2735,\\
Beijing 100190, China \\
Department of Applied Physics, Xi' an Jiaotong University\\ Xi' an
710049, China\\ huaifan.li@stu.xjtu.edu.cn}

\author{Bin Hu}

\address{Key Laboratory of Frontiers in Theoretical Physics, Institute of
Theoretical Physics, Chinese Academy of Sciences, P.O. Box 2735,\\
Beijing 100190, China \\ hubin@itp.ac.cn}

\maketitle

\pub{Received (Day Month Year)}{Revised (Day Month Year)}

\begin{abstract}
We established the equivalence between the local Hawking temperature
measured by the time-like Killing observer located at some positions
$r$ with finite distances from the outer horizon $r_+$ in the
5-dimensional spinning black hole space with both negative and
positive constant curvature, and the Unruh temperature measured by
the Rindler observer with constant acceleration in the 6-dimensional
flat space by employing the globally embedding approach.

\keywords{Black hole; Hawking temperature; Unruh temperature;
Globally embedding approach.}
\end{abstract}

\ccode{PACS Nos.: 04.70.Dy; 04.50.Gh}

\newcommand{\no}{\nonumber}

\def\be{\begin{equation}}
\def\ee{\end{equation}}
\def\bea{\begin{eqnarray}}
\def\eea{\end{eqnarray}}

\newcommand{\ha}{{1 \over 2}}
\newcommand{\la}{\lambda}
\newcommand{\rc}{\nonumber\\}

\newcommand{\bear}{\begin{eqnarray}}

\newcommand{\eear}{\end{eqnarray}}
\newcommand{\de}{\partial}
\newbox\pippobox

\def\be{\begin{equation}}
\def\ee{\end{equation}}
\def\bea{\begin{eqnarray}}
\def\eea{\end{eqnarray}}
\def\bx{{\bf x}}
\def\by{{\bf y}}
\def\br{{\bf r}}
\def\bk{{\bf k}}
\def\bp{{\bf p}}
\def\bn{{\bf n}}
\def\eb{{\bf e}}
\def\bB{{\bf B}}
\def\a{\alpha}
\def\e{\epsilon}
\def\m{\mu}
\def\n{\nu}
\def\9{\nabla}
\def\r{\rho}
\def\s{\sigma}
\def\vs{\varsigma}
\def\t{\theta}
\def\T{\Theta}
\def\g{\gamma}
\def\z{\zeta}
\def\b{\beta}
\def\o{\omega}
\def\lam{\lambda}
\def\y{\psi}
\def\k{\chi}
\def\kp{\kappa}
\def\h{\eta}
\def\d{\delta}
\def\dd{{\rm d}}
\def\D{\Delta}
\def\p{\phi}
\def\P{\Phi}
\def\N{{\cal N}}
\def\R{R_{\n}}
\def\nn{\nonumber}
\def\half{\frac12}
\def\le{\left}
\def\ri{\right}
\def\6{\partial}
\def\bbox{\nabla^2}
\def\na{\nabla}
\def\vp{\varphi}
\def\ve{\varepsilon}
\def\f{\frac}
\def\ma{\mathcal}
\def\la{\langle}
\def\ra{\rangle}
\def\B{\Big}
\def\Mp{M_{pl}}
\def\tld{\tilde}
\def\0{(0)}
\def\half{\f{1}{2}}
\def\>{\rightarrow}

\section{Introduction}
In the past decade it was shown that, for both Hawking
\cite{Hawking:1974sw,Bekenstein:1973ur,Gibbons:1977mu} and Unruh
\cite{Davies,Unruh} effects, temperature and entropy emerge from
information loss associated with real and accelerated-observer
horizons, respectively. It was firstly demonstrated in \cite{Narn}
that an observer with a constant acceleration $a$ in de Sitter space
will detect a temperature given by $\sqrt{a^2+1/l^2}/2\pi$, where
$l$ is the de Sitter radius. This result was soon generalized by
Deser and Levin \cite{Deser:1997ri,Deser:1998bb} that the local
Hawking temperatures $\sqrt{a^2\pm1/l^2}/2\pi$ measured by
accelerated detectors in (anti-)de Sitter (AdS/dS) geometries can be
obtained from their corresponding Rindler observors with constant
accelerations by using the Global Embedding Approach (GEA). For more
examples on the equivalence, such as Banados-Teitelboim-Zanelli,
Schwarzschild, Schwarzschild-AdS (dS), and Reissner-Nordstr\" om
solutions into higher dimensional Minkowskian spaces etc., see
\cite{Deser:1998xb,Kim,Chen:2004qw,Chen:2005wj} and references
therein.

On the other hand, the so-called BTZ (Banados-Teitelboim-Zanelli)
black hole solutions  \cite{BTZ,BHTZ} have played the important role
in understanding microscopic degrees of freedom of black hole. The
BTZ black hole, which is constructed by identifying points along the
orbit of a Killing vector in a three dimensional anti-de Sitter
space, is an exact solution of Einstein field equations with a
negative cosmological constant in three dimensions. This kind of
black hole has a topology of $\ma{M}_2\times S^1$, where $\ma{M}_2$
denotes a conformal Minkowski space in two dimensions. Following the
same logic, one can construct analogues of the BTZ solution, the
so-called constant curvature (CC) black holes in higher $n\geq4$
dimensional AdS spaces \cite{Holst,Banados:1997df,Bana2}. Because
$n$-dimensional anti-de Sitter space has the topology $R_{n-1}\times
S^{1}$, thel CC black holes have topology of $\ma{M}_{n-1}\times
S^1$, which is quite different from the known topology of
$\ma{M}_2\times S^{n-2}$ for the usual black holes in $n$
dimensions. In addition, the exterior region  of these CC black
holes is time-dependent and thus, there is no global time-like
Killing vector~\cite{Holst}. Because of  this feature, it is
difficult to discuss Hawking radiation and thermodynamics associated
with these black holes. For example, see \cite{Mann,Ross,Hut} and
references therein.

Comparing with the anti-de Sitter case, the $n$-dimensional de
Sitter space has the different topology $R_1\times S^{n-1}$. Similar
to the negative constant curvature case,  a positive constant
curvature spacetime was constructed in \cite{Cai:2002yy} by
identifying points along a rotation Killing vector in a de Sitter
space. Such solutions turn out to be counterparts of the
three-dimensional Schwarzschild de Sitter solution in higher
dimensions, and have an associated cosmological horizon. There is a
parameter in the solution, which can be explained as the size of
cosmological horizon.

The equivalence between the lower dimensional Hawking temperature
and higher dimensional Unruh temperature has been well established
for the general spinning BTZ black hole in
\cite{Deser:1998xb,Russo:2008gb} and the spinnless negative/positive
CC black hole in \cite{Cai:2010bv}. It should be noted that the
Hawking temperature obtained by Cai and Myung \cite{Cai:2010bv} is
the same as that obtained from semi-classical tunneling method
\cite{Yale:2010qq}. Recently, a reduced approach to the study of
Hawking/Unruh effects including their unification was proposed by
authors of \cite{Banerjee:2010ma}. The primary goal of this work is
to generalize the equivalence for the spinnless CC black hole
established in \cite{Cai:2010bv} to the spinning one. The rest of
this paper is organized as follows. In section II, we briefly review
the construction of the spinless negative/positive CC black hole
from the (anti)-de Sitter space by identifying the points along a
boost direction. In section III and IV, we firstly construct the
5-dimensional spinning black hole with negative and positive
constant curvature, respectively, and then establish the equivalence
between the Hawking temperature in 5-dimension and the Unruh
temperature in 6-dimension. Finally, we conclude in section V.

\section{review on the spinless constant curvature black hole}
In this section, we begin with a brief review the construction of
the spinless CC black hole from the (anti-)de Sitter space by
identifying the points along one boost direction. The
$n$-dimensional (anti-)de Sitter space is defined as (the universal
covering) of the hypersurface embedded into a $(n+1)$-dimensional
Minkowskian space, satisfying
 \be\label{embd}
 -x_0^2+x_1^2+\cdots +x_{n-1}^2\mp x_n^2=\mp l^2\;,\ee
here and in the following paragraph of this section, we denote the
upper and lower sign for anti-de Sitter and de Sitter space,
respectively. The (anti-)de Sitter space adimits the boost along
Killing vector $\xi=(r_+/l)(x_{n-1}\6_n\pm x_n\6_{n-1})$ with norm
$\xi^2=(r_+^2/l^2)(\mp x_{n-1}^2+x_n^2)$.

In what follows, we will firstly discuss the anti-de Sitter case, in
which the $(n+1)$-dimensional flat space has two timelike
coordinates
 \be\label{mink}
 ds^2=-dx_0^2+dx_1^2+\cdots +dx_{n-1}^2-dx_n^2\;.\ee

To go further in the discussion, let us introduce the local Kruskal
coordinates ($y_{\a}, \phi$) on anti-de Sitter space (in the region
$\xi^2>0$),
 \bea
 x_{\a}&=&\f{2ly_{\a}}{1-y^2}\;,\qquad \a=0,\cdots,n-2\;,\\
 x_{n-1}&=&\f{lr}{r_+}\sinh\le(\f{r_+\phi}{l}\ri)\;,\\
 x_n&=&\f{lr}{r_+}\cosh\le(\f{r_+\phi}{l}\ri)\;
 \eea
where $r$ and $y^2$ are defined as
 \be\label{ry}
 r=r_+\f{1+y^2}{1-y^2}\;,\ee
and $y^2=\h_{\a\b}y^{\a}y^{\b}$ $\le[\h_{\a\b}=\rm{diag}(-1,1,\cdots
,1)\ri]$, respectively. The coordinate ranges are
$-\infty<\phi<\infty$ and $-\infty<y^{\a}<\infty$ with the
restriction $-1<y^2<1$. By using Kruskal coordinate, the induced
metric on the anti-de Sitter space reads
 \be\label{kruskal}
 ds^2=\f{l^2(r+r_+)^2}{r_+^2}dy^{\a}dy^{\b}\h_{\a\b}+r^2d\phi^2\;.\ee
Without lossing any information, we shall restrict the discussion to
the five dimensional case. The negative CC black hole can be easily
obtained by introducing the local ``spherical'' coordinates
$(t,r,\theta,\chi)$
 \bea\label{schwatzlike}
 y_0&=&f\cos\theta\sinh(r_+t/l)\;,\qquad y_2=f\sin\theta\sin\chi\;,\\
 y_1&=&f\cos\theta\cosh(r_+t/l)\;,\qquad y_3=f\sin\theta\cos\chi\;,\eea
with $f(r)=[(r-r_+)/(r+r_+)]^{1/2}$. (Note that these coordinates,
with ranges $0<\theta<\pi/2$, $0\leq\chi<2\pi$, $-\infty<t<\infty$,
and $r_+<r<\infty$, do not cover the whole region $r>r_+$ but only
$-1<y_2,y_3<1$.) Using these new coordinates, the metric
(\ref{kruskal}) can be casted into the Schwazschild form
 \be\label{schwatzlikemetric}
 ds^2=l^2N^2d\Omega_3^2+N^{-2}dr^2+r^2d\phi^2\;,\ee
with $N^2(r)=(r^2-r_+^2)/l^2$ and
 \be
 d\Omega_3^2=-\cos^2\theta dt^2+\f{l^2}{r_+^2}(d\theta^2+\sin^2\theta d\chi^2)\;.\ee

In this coordinate frame, the Killing vector, generated a boost,
becomes into $\xi=\6_{\phi}$ with norm $\xi^2=r^2$, and the spinless
negative CC black hole can be obtained by identifying the points
along the Killing vector $\xi$
 \be
 \phi\sim \phi+2n\pi\;,\qquad n\in \mathbb Z\;.\ee
The horizon for the negative CC black hole in these coordinates is
located at $r=r_+$, the point where $N^2$ vanishes.

The positive CC black hole can be constructed analogously, by
identifying the points along the same Killing vector $\xi=\6_{\phi}$
with norm $\xi^2=r^2$. In details, we firstly embed the
$n$-dimensional de Sitter space into  a $(n+1)$-dimensional
Minkowskian space
  \be\label{mink2}
 ds^2=-dx_0^2+dx_1^2+\cdots +dx_{n-1}^2+dx_n^2\;.\ee
Furthermore, we define the Kruskal coordinates $(y_{\a},\phi)$ on
the $n$-dimensional de Sitter space in the region $(0\leq \xi^2\leq
r_+^2)$,
 \bea
 x_{\a}&=&\f{2ly_{\a}}{1-y^2}\;,\qquad \a=0,\cdots,n-2\;,\\
 x_{n-1}&=&\f{lr}{r_+}\sin\le(\f{r_+\phi}{l}\ri)\;,\\
 x_n&=&\f{lr}{r_+}\cos\le(\f{r_+\phi}{l}\ri)\;,
 \eea
with
 \bea\label{ry2}
 r=r_+\f{1-y^2}{1+y^2}\;,\qquad y^2=\h_{\a\b}y^{\a}y^{\b}\;,\no \\  \quad \h_{\a\b}=\rm{diag}(-1,1,\cdots
 ,1)\;.\eea
Here the coordinate range is $-\infty<y_{\a}<+\infty$, and
$-\infty<\phi<+\infty$ with the restriction $-1<y^2<1$ in order to
keep $r$ positive. Under the Kruskal coordinate frame, the induced
metric takes the same from with (\ref{kruskal}). We can also
introduce Schwarzschild coordinates to describe the solution. Using
local spherical coordinates $(t,r,\theta,\xi)$ defined as
(\ref{schwatzlike}) with  $f(r)=[(r_+-r)/(r+r_+)]^{1/2}$, and the
coordinate ranges are $0<\theta<\pi/2$, $0\leq\chi<2\pi$,
$-\infty<t<\infty$, and $0<r<r_+$, we find that the solution can be
expressed as \be\label{schwatzlikemetric2}
 ds^2=l^2N^2d\Omega_3^2+N^{-2}dr^2+r^2d\phi^2\;,\ee
with $N^2(r)=(r_+^2-r^2)/l^2$ and
 \be
 d\Omega_3^2=-\cos^2\theta dt^2+\f{l^2}{r_+^2}(d\theta^2+\sin^2\theta d\chi^2)\;.\ee
Like what happens for the negative constant curvature case, the
spinless positive CC black hole can be constructed by identifying
the points along the Killing vector $\xi=\6_{\phi}$ with norm
$\xi^2=r^2$
 \be
 \phi\sim \phi+2n\pi\;,\qquad n\in \mathbb Z\;.\ee
In these coordinates $r=r_+$ is the cosmological horizon. The only
difference is that $N^2=(r^2-r_+^2)/l^2$ in the negative CC black
hole space there is replaced by $N^2=(r_+^2-r^2)/l^2$ in the
positive one.

\section{\label{negative ccbh} spinning negative constant curvature black hole}
In this section, we firstly construct the 5-dimensional spinning
negative CC black hole by virtue of the GEA, then we calculate both
the local Hawking temperature observed by a time-like Killing
observer in the 5-dimensional hypersurface and the corresponding
Unruh temperature observed by a Rindler observer in the
6-dimensional flat space.

By using the ``spherical'' coordinate $(t,\rho,\theta,\phi,\chi)$
 \bea
 x_0=l\sinh\rho\cos\theta\sinh\le(\f{r_+}{l}t-\f{r_-}{l}\phi\ri)\;,\no \\
 x_1=l\sinh\rho\cos\theta\cosh\le(\f{r_+}{l}t-\f{r_-}{l}\phi\ri)\;,\no \\
 x_2=l\sinh\rho\sin\theta\sin\chi\;, \\
 x_3=l\sinh\rho\sin\theta\cos\chi\;,\no \\
 x_4=l\cosh\rho\sinh\le(\f{r_+}{l}\phi-\f{r_-}{l}t\ri)\;,\no \\
 x_5=l\cosh\rho\cosh\le(\f{r_+}{l}\phi-\f{r_-}{l}t\ri)\;,\no \eea
with $r_+$ and $r_-$ ($r_+>r_-$) two arbitrary real constants with
dimensions of length, the 5-dimensional spinning negative CC black
hole can be obtained
 \bea\label{rotmetric}
 ds^2&=&\cos^2\theta\le[-N^2l^2dt^2+r^2(d\phi +N^{\phi}dt)^2\ri] \no \\
 &&+N^{-2}dr^2+l^2\f{r^2-r_+^2}{r_+^2-r_-^2}\le(d\theta^2+\sin ^2\theta d\chi^2\ri)\\
 &&+l^2\f{r^2-r_-^2}{r_+^2-r_-^2}\sin ^2\theta \le(\f{r_+}{l}d\phi-\f{r_-}{l}dt\ri)^2\;, \no  \eea
by identifying
 \be\label{indentify}
 \phi\sim \phi+2n\pi\;,\qquad n\in \mathbb Z\;,\ee
where
 \bea
 N^2&=&\f{(r^2-r_+^2)(r^2-r_-^2)}{l^2r^2}\;,\\
 N^{\phi}&=&-\f{r_+r_-}{r^2}\;,\eea
and
 \be
 r^2=r_+^2\cosh^2\rho-r_-^2\sinh^2\rho\;.\ee
In these coordinates, $-\infty<t<\infty$, $0<\theta<\pi/2$,
$0\leq\chi<2\pi$ and $r_+<r<\infty$.

In the 5-dimensional spinning space, the local observer is chosen as
the one whose trajectory follows the time-like Killing vector
$\z=\6_t+r_-/r_+\6_{\phi}$, namely the ``time-like Killing
observer'' $(\ma{O}_K)$
 \be\label{obs}
 \phi=\f{r_-}{r_+}t\;,\qquad r, \theta, \chi={\rm const.}\;,\ee
then the proper velocity of $(\ma{O}_K)$ is
 \bea\label{vel}
u^{\m}=(t,r,\theta,\phi,\chi)=
 \le(\f{r_+}{\cos\theta\sqrt{(r^2-r_+^2)(r^2-r_-^2)}},0,0,\f{r_-}{\cos\theta\sqrt{(r^2-r_+^2)(r^2-r_-^2)}},0\ri)\;,\eea
and the proper acceleration for $(\ma{O}_K)$ reads
 \be\label{a5}
 a_5^{\mu}=\le(0,\f{r^2-r_-^2}{l^2r},-\f{(r_+^2-r_-^2)\tan\theta}{l^2(r^2-r_+^2)},0,0\ri)\;.\ee
Thus the $a_5$ equals
 \be\label{a52}
 a_5=\f{\sqrt{(r^2-r_-^2)+\tan^2\theta(r_+^2-r_-^2)}}{l\sqrt{r^2-r_+^2}}\;.\ee
The Killing vector tangent to the worldline of the time-like
observer ($\ma{O}_K$) is $\z=\6_t+r_-/r_+\6_{\phi}$, consequently,
the Hawking temperature which enters the thermodynamical relations
can be obtained through the surface gravity
[$\kp^2=-\half(\na^{\m}\xi^{\n})(\na_{\m}\xi_{\n})$]
 \be\label{surf g2}
 2\pi T_{\rm{HK}}=\kp=\f{r_+^2-r_-^2}{r_+l}\;.\ee
This result is consistent with the one which is obtained by
embedding the spinning CC black hole into a Chern-Simons
supergravity theory \cite{Banados:1997df}. However, as pointed out
in \cite{Cai:2010bv} the method used in \cite{Banados:1997df} has
two drawbacks, one is that the result cannot be degenerated to the
spinnless case, the other is that it cannot be generalized to other
dimensions. Along the integral curves generated by the Killing
vector $\z$ the line element becomes
 \bea\label{metric2}
 ds^2&=&\cos^2\theta\le[-N^2l^2dt^2+r^2(\f{r_-}{r_+}dt +N^{\phi}dt)^2\ri]+\cdots\;,\nn\\
 &=&-\cos^2\theta\f{(r^2-r_+^2)(r_+^2-r_-^2)}{r_+^2}dt^2+\cdots\;.\eea
Armed with the above results, we can define the Tolman temperature
which is the local Hawking temperature observed by the time-like
Killing observer ($\ma{O}_{K}$) located at $r$
 \bea\label{hawking T}
 2\pi T=\f{\kp}{\sqrt{-\hat g_{00}}}=\f{\sqrt{r_+^2-r_-^2}}{l\cos\theta\sqrt{r^2-r_+^2}}\;,\eea
where $\hat g_{00}$ is the Tolman redshift factor which can be read
off from (\ref{metric2}).

On the other hand, the worldline of the time-like Killing observer
in the 5-dimensional spinning black hole space coincides with the
trajectory of Rindler observer with the constant acceleration
 \be\label{a6}
 a_6^{-2}=x_1^2-x_0^2=l^2\f{r^2-r_+^2}{r_+^2-r_-^2}\cos^2\theta\;,\ee
thus, the corresponding Unruh temperature reads
 \be\label{unruhT}
 2\pi T_{DU}=a_6=\f{\sqrt{r_+^2-r_-^2}}{l\cos\theta\sqrt{r^2-r_+^2}}=\sqrt{\f{1}{l^2}+a_5^2}\;.\ee
From (\ref{hawking T}) and (\ref{unruhT}) we can easily see that the
local Hawking temperature measured by a time-like Killing observer
in the 5-dimensional space is nothing but the Unruh temperature
observed by a Rindler observer in the 6-dimensional flat space.

\section{\label{positive ccbh} spinning positive constant curvature black hole}
Analogously, the 5-dimensional spinning positive CC black hole can
also be constructed by using the ``spherical'' coordinate
$(t,\rho,\theta,\phi,\chi)$
 \bea \label{sph1}
 x_0=l\sin\rho\cos\theta\sinh\le(\f{r_+}{l}t-\f{r_-}{l}\phi\ri)\;,\no \\ \qquad
 x_1=l\sin\rho\cos\theta\cosh\le(\f{r_+}{l}t-\f{r_-}{l}\phi\ri)\;, \no \\ \qquad
 x_2=l\sin\rho\sin\theta\sin\chi\;,\\\qquad
 x_3=l\sin\rho\sin\theta\cos\chi\;,\no \\\qquad
 x_4=l\cos\rho\sin\le(\f{r_+}{l}\phi+\f{r_-}{l}t\ri)\;,\no  \\
 x_5=l\cos\rho\cos\le(\f{r_+}{l}\phi+\f{r_-}{l}t\ri)\;. \no \eea
Plugging (\ref{sph1}) into the 6-dimensional Minkowskian metric and
identifying $\phi\sim\phi+2n\pi$ with $n\in \mathbb Z$, one can
derive the induced metric which describes a 5-dimensional spinning
black hole with positive constant curvature
 \bea\label{protmetric}
 ds^2&=&\cos^2\theta\le[-N^2l^2dt^2+r^2(d\phi +N^{\phi}dt)^2\ri]+N^{-2}dr^2\no \\
 &&+l^2\f{r_+^2-r^2}{r_+^2+r_-^2}\le(d\theta^2+\sin ^2\theta d\chi^2\ri)+l^2\f{r^2+r_-^2}{r_+^2+r_-^2}\sin ^2\theta \le(\f{r_+}{l}d\phi+\f{r_-}{l}dt\ri)^2\;,\eea
where
 \bea
 N^2&=&\f{(r_+^2-r^2)(r^2+r_-^2)}{l^2r^2}\;,\\
 N^{\phi}&=&\f{r_+r_-}{r^2}\;,\qquad (0<r_-<r_+)\;,\eea
and
 \be
 r^2=r_+^2\cos^2\rho-r_-^2\sin^2\rho\;,\qquad 0<r<r_+\;.\ee
The ranges of other ``spherical'' coordinates ($t,\theta,\phi,\chi$)
are the same as the case with negative constant curvature. In these
coordinates, the cosmological horizon locates at $r_+$ and the
ranges $0<r<r_+$ represents the interior of the horizon
\cite{Park:1998qk}.

The 5D CC black hole is constructed by indentifying
 \be\label{pindentify}
 \phi\sim \phi+2n\pi\;,\qquad n\in \mathbb Z\;\ee
The time-like Killing observer $(\ma{O}_K)$ is chosen as
 \be\label{pobs}
 \phi=-\f{r_-}{r_+}t\;,\qquad r, \theta, \chi={\rm const.}\;,\ee
then the corresponding proper velocity is
 \be\label{pvel}
 u^{\m}=(t,r,\theta,\phi,\chi)=\le(\f{r_+}{\cos\theta\sqrt{(r_+^2-r^2)(r_+^2+r_-^2)}},0,0,\f{-r_-}{\cos\theta\sqrt{(r_+^2-r^2)(r_+^2+r_-^2)}},0\ri)\;.\ee
The proper acceleration for observer $(\ma{O}_K)$ can be derived by
differentiating the proper velocity (\ref{pvel}) with respect to the
proper time
 \be\label{pa5}
 a_5^{\mu}=\le(0,-\f{r^2+r_-^2}{l^2r},-\f{(r_+^2+r_-^2)\tan\theta}{l^2(r_+^2-r^2)},0,0\ri)\;,\ee
thus the $a_5$ equals
 \be\label{pa52}
 a_5=\f{\sqrt{(r^2+r_-^2)+\tan^2\theta(r_+^2+r_-^2)}}{l\sqrt{r_+^2-r^2}}\;.\ee
The Killing vector tangent to the worldline of the time-like
observer ($\ma{O}_K$) reads $\z=\6_t-r_-/r_+\6_{\phi}$, so the
Hawking temperature reads
 \be\label{psurf g2}
 2\pi T_{{\rm HK}}=\kp=\f{r_+^2+r_-^2}{r_+l}\;.\ee
This result is consistent with those obtained in \cite{Park:1998qk}.
Along the integral curves generated by the Killing vector $\z$, the
line element becomes
 \bea\label{pmetric2}
 ds^2&=&\cos^2\theta\le[-N^2l^2dt^2+r^2(\f{r_-}{r_+}dt +N^{\phi}dt)^2\ri]+\cdots\;,\nn\\
 &=&-\cos^2\theta\f{(r_+^2-r^2)(r_+^2+r_-^2)}{r_+^2}dt^2+\cdots\;.\eea
The local Hawking temperature measured by the time-like Killing
observer ($\ma{O}_{K}$) located at $r$ equals the Hawking
temperature in (\ref{psurf g2}) rescaled by a redshift factor
 \bea\label{phawking T}
 2\pi T=\f{\kp}{\sqrt{-\hat g_{00}}}=\f{\sqrt{r_+^2+r_-^2}}{l\cos\theta\sqrt{r_+^2-r^2}}\;,\eea
where the redshift factor are derived in (\ref{pmetric2}).

The corresponding Unruh acceleration and temperature can also be
obtained by using the same logic in the negative case
 \be\label{pa6}
 a_6^{-2}=x_1^2-x_0^2=l^2\f{r_+^2-r^2}{r_+^2+r_-^2}\cos^2\theta\;,\ee
and
 \be\label{punruhT}
 2\pi T_{DU}=a_6=\f{\sqrt{r_+^2+r_-^2}}{l\cos\theta\sqrt{r_+^2-r^2}}=\sqrt{\f{1}{l^2}+a_5^2}\;.\ee
As we expect that the tempratures (\ref{phawking T}) and
(\ref{punruhT}) coincide exactly.

\section{\label{con}Conclusion}
In this work, we established the equivalence between the local
Hawking temperature measured by the time-like Killing observer
located at some positions $r$ with finite distances from the outer
horizon $r_+$ in the 5-dimensional spinning CC black hole space, and
the Unruh temperature measured by the Rindler observer with constant
acceleration in the 6-dimensional flat space by employing the
globally embedding approach. For the spinning black hole with
negative constant curvature, the local Hawking or Unruh temperature
equals to $\sqrt{r_+^2-r_-^2}/2\pi l\cos\theta\sqrt{r^2-r_+^2}$,
while for the postive case, the temperature reads
$\sqrt{r_+^2+r_-^2}/2\pi l\cos\theta\sqrt{r_+^2-r^2}$, where $l$ is
the radius of (anti-)de Sitter spaces and $\theta$ is the spherical
coordinate of time-like Killing observer. Our results can recover
the previous one in both the general spinning BTZ black hole
\cite{Bana2} when one sets $\theta=0$ and the spinnless black hole
with negative/positive constant curvature \cite{Cai:2010bv} when one
sets both $\theta=0$ and $r_-=0$.

\section{Acknowledgments}
We are grateful to Hai-Qing Zhang, Zhang-Yu Nie and Yun-Long Zhang
for their various discussions. We especially thank Rong-Gen Cai for
the useful comments on draft and the persistent encouragements
during all stages of this work. BH thanks the nice accommodation
during ``the 5th Asian Winter School on Strings, Particles and
Cosmology'', Jeju, Korea. This work was supported in part by a grant
from Chinese Academy of Sciences and in part by the National Natural
Science Foundation of China under Grant Nos. 10821504, 10975168,
11035008, 11075098 and 11175109 and by the Ministry of Science and
Technology of China under Grant No. 2010CB833004.



\vspace*{0.2cm}
\begin{thebibliography}{99}

\bibitem{Hawking:1974sw}
  S.~W.~Hawking,
  {\em ``Particle Creation By Black Holes,''}
  Commun.\ Math.\ Phys.\  {\bf 43}, 199 (1975)
  [Erratum-ibid.\  {\bf 46}, 206 (1976)].
\bibitem{Bekenstein:1973ur}
  J.~D.~Bekenstein,
  {\em ``Black holes and entropy,''}
  Phys.\ Rev.\  D {\bf 7}, 2333 (1973).
\bibitem{Gibbons:1977mu}
  G.~W.~Gibbons and S.~W.~Hawking,
  {\em ``Cosmological Event Horizons, Thermodynamics, And Particle Creation,''}
  Phys.\ Rev.\  D {\bf 15}, 2738 (1977).


\bibitem{Davies} P.~C.~W.~Davies,
  {\em ``Scalar particle production in Schwarzschild and Rindler metrics,''}
  J.\ Phys.\ A  {\bf 8}, 609 (1975).
\bibitem{Unruh} W.~G.~Unruh,
  {\em ``Notes on black hole evaporation,''}
  Phys.\ Rev.\  D {\bf 14}, 870 (1976).

\bibitem{Narn}H.~Narnhofer, I.~Peter and W.~E.~Thirring,
  {\em ``How hot is the de Sitter space?,''}
  Int.\ J.\ Mod.\ Phys.\  B {\bf 10}, 1507 (1996).


\bibitem{Deser:1997ri}
  S.~Deser and O.~Levin,
  {\em ``Accelerated detectors and temperature in (anti) de Sitter spaces,''}
  Class.\ Quant.\ Grav.\  {\bf 14}, L163 (1997)
  [arXiv:gr-qc/9706018].
\bibitem{Deser:1998bb}
  S.~Deser and O.~Levin,
  {\em ``Equivalence of Hawking and Unruh temperatures through flat space
  embeddings,''}
  Class.\ Quant.\ Grav.\  {\bf 15}, L85 (1998)
  [arXiv:hep-th/9806223].

\bibitem{Deser:1998xb}
  S.~Deser and O.~Levin,
  {\em ``Mapping Hawking into Unruh thermal properties,''}
  Phys.\ Rev.\  D {\bf 59}, 064004 (1999)
  [arXiv:hep-th/9809159].

\bibitem{Kim}Y.~W.~Kim, J.~Choi and Y.~J.~Park,
  {\em ``Local free-fall temperature of a RN-AdS black hole,''}
  Int.\ J.\ Mod.\ Phys.\  A {\bf 25}, 3107 (2010)
  [arXiv:0909.3176 [gr-qc]].

\bibitem{Chen:2004qw}
  H.~Z.~Chen, Y.~Tian, Y.~H.~Gao and X.~C.~Song,
  {\em ``The GEMS Approach to Stationary Motions in the Spherically Symmetric
  Spacetimes,''}
  JHEP {\bf 0410}, 011 (2004)
  [arXiv:gr-qc/0409107].

\bibitem{Chen:2005wj}
  H.~Z.~Chen and Y.~Tian,
  {\em ``Note on the generalization of the global embedding Minkowski spacetime
  approach,''}
  Phys.\ Rev.\  D {\bf 71}, 104008 (2005).

\bibitem{BTZ} M.~Banados, C.~Teitelboim and J.~Zanelli,
  {\em ``The Black hole in three-dimensional space-time,''}
  Phys.\ Rev.\ Lett.\  {\bf 69}, 1849 (1992)
  [arXiv:hep-th/9204099];
\bibitem{BHTZ}
 M.~Banados, M.~Henneaux, C.~Teitelboim and J.~Zanelli,
  {\em ``Geometry of the (2+1) black hole,''}
  Phys.\ Rev.\  D {\bf 48}, 1506 (1993)
  [arXiv:gr-qc/9302012].

\bibitem{Holst}S.~Holst and P.~Peldan,
{\em ``Black holes and causal structure in anti-de Sitter isometric
spacetimes,''} Class.\ Quant.\ Grav.\  {\bf 14}, 3433 (1997)
[arXiv:gr-qc/9705067].

\bibitem{Banados:1997df}
  M.~Banados,
  {\em ``Constant curvature black holes,''}
  Phys.\ Rev.\  D {\bf 57}, 1068 (1998)
  [arXiv:gr-qc/9703040].

\bibitem{Bana2}M.~Banados, A.~Gomberoff and C.~Martinez,
{\em ``Anti-de Sitter space and black holes,''} Class.\ Quant.\
Grav.\  {\bf 15}, 3575 (1998) [arXiv:hep-th/9805087].

\bibitem{Mann}J.~D.~Creighton and R.~B.~Mann,
{\em ``Entropy of CC black holes in general relativity,''} Phys.\
Rev.\ D {\bf 58}, 024013 (1998) [arXiv:gr-qc/9710042].

\bibitem{Ross}  S.~F.~Ross and G.~Titchener,
{\em ``Time-dependent spacetimes in AdS/CFT: Bubble and black
hole,''}
  JHEP {\bf 0502}, 021 (2005)
  [arXiv:hep-th/0411128].

\bibitem{Hut}J.~A.~Hutasoit, S.~P.~Kumar and J.~Rafferty,
{\em ``Real time response on $dS_3$: the Topological AdS Black Hole
and the Bubble,''}
  JHEP {\bf 0904}, 063 (2009)
  [arXiv:0902.1658 [hep-th]].

\bibitem{Cai:2002yy}
  R.~G.~Cai,
  {\em ``A positive constant curvature space and associated entropy,''}
  Phys.\ Lett.\  B {\bf 552}, 66 (2003)
  [arXiv:hep-th/0207053].

\bibitem{Cai:2010bv}
  R.~G.~Cai and Y.~S.~Myung,
  {\em ``Hawking temperature for constant curvature black bole and its analogue in de Sitter space,''}
  arXiv:1012.5709 [hep-th].

\bibitem{Yale:2010qq}
  A.~Yale,
  {\em ``Thermodynamics of Constant Curvature Black Holes Through Semi-Classical
  Tunneling,''}
  arXiv:1012.2114 [gr-qc].

\bibitem{Park:1998qk}
  M.~I.~Park,
  {\em ``Statistical entropy of three-dimensional Kerr-de Sitter space,''}
  Phys.\ Lett.\  B {\bf 440}, 275 (1998)
  [arXiv:hep-th/9806119].

\bibitem{Banerjee:2010ma}
  R.~Banerjee and B.~R.~Majhi,
  {\em ``A New Global Embedding Approach to Study Hawking and Unruh Effects,''}
  Phys.\ Lett.\  B {\bf 690}, 83 (2010)
  [arXiv:1002.0985 [gr-qc]].

\bibitem{Russo:2008gb}
  J.~G.~Russo and P.~K.~Townsend,
  {\em ``Accelerating Branes and Brane Temperature,''}
  Class.\ Quant.\ Grav.\  {\bf 25}, 175017 (2008)
  [arXiv:0805.3488 [hep-th]].
\end{thebibliography}
\end{document}